\documentclass[12pt]{article}
\usepackage{amssymb,latexsym,epsfig,cite}

\def\be{\begin{equation}}
\def\ee{\end{equation}}
\def\bea{\begin{eqnarray}}
\def\eea{\end{eqnarray}}
\begin{document}

\baselineskip=18pt
\hfill

\begin{center}
{\LARGE{\bf The Phenomenon of Darboux}}

\smallskip
{\LARGE{\bf Displacements}}

\vskip1cm

David J. Fern\'andez C.${}^1$, Bogdan Mielnik${}^{1,2}$ \\
Oscar Rosas--Ortiz${}^1$ and Boris F. Samsonov${}^3$
\vskip0.5cm

{\it
${}^1$Departamento de F\'{\i}sica, CINVESTAV \\ 
AP 14-740, 07000 M\'exico DF, Mexico
\\  [1ex]
${}^2$Institute of Theoretical Physics, Warsaw University \\ Ho\.{z}a 69
Warsaw, Poland
\\  [1ex]
${}^3$Department of Quantum Field Theory, Tomsk State
University \\ 36 Lenin Ave., 634050 Tomsk, Russia
}
\end{center}
\vskip0.5cm

\begin{abstract}
\baselineskip=15pt
For a class of Schr\"odinger Hamiltonians the supersymmetry
transformations can degenerate to simple coordinate displacements.  We
examine this phenomenon and show that it distinguishes the Weierstrass
potentials including the one-soliton wells and periodic Lam\'e functions.
A supersymmetric sense of the addition formula for the Weierstrass
functions is elucidated. 
\end{abstract}

\vfill 



\smallskip

\begin{description}
\item[Key-Words:] Supersymmetry, factorization, addition laws
\item[PACS:]  03.65.Ge, 03.65.Fd, 03.65.Ca
\end{description}

\newpage

\baselineskip=15pt

\section{Introduction}

The Darboux (supersymmetry)  transformations \cite{Darb} are an efficient
tool to extend the class of exactly solvable spectral problems for the
Schr\"odinger's Hamiltonians in one space dimension
\cite{Schro,Infe,MMN,Andr,bo88,lr91,sta95}, permitting to see that new
classes of exact solutions are not an exclusive achievement of General
Relativity \cite{Krasi}. The use of the Darboux method turns of special
interest in particle and statistical physics, classical mechanics,
mathematical physics, biological systems and other areas
\cite{wi81,Zakha,bs97,dfr97} (see also the monographs
\cite{ms91,ju96,ba01,cooper}). If the initial Schr\"odinger's Hamiltonian in
$L^2 ({\mathbb R})$ is: 

\be
H =-\frac{1}{2} \frac{d^2}{dx^2} + V(x)
\label{scr1}
\ee
the original (1-st order) Darboux algorithm \cite{Darb,Schro,Infe} is
equivalent to the operation of `transporting' through $H$ a 1-st order
differential operator $A$ or its adjoint $A^{\dagger}$: 
\be
A = \frac{1}{\sqrt{2}}\left[\frac{d}{dx} + \alpha(x)\right], \qquad
A^{\dagger} = \frac{1}{\sqrt{2}}\left[-\frac{d}{dx} +
\alpha(x)\right]
\label{intertwiners}
\ee
leading to a new Hamiltonian $\tilde H$
\be
A H = \tilde H A   \Leftrightarrow   H A^{\dagger} = A^{\dagger}
\tilde H \label{intertwining}
\ee
again of Schr\"odinger's type
\be
\tilde H =-\frac{1}{2} \frac{d^2}{dx^2} + \tilde V(x)
\label{scr2}
\ee
provided that the function $\alpha (x)$ (the {\it superpotential}) 
fulfills the integrability condition in form of the Riccati equation: 
\be
-\alpha'(x) + \alpha^2 (x) = 2[V(x)-\epsilon],
\label{riccati1}
\ee
where $\epsilon$ is a {\it factorization constant} measured in energy
units though not necessarily belonging to the spectrum of $H$ \cite{more}.
If (\ref{intertwining}-\ref{riccati1}) hold, the new potential $\tilde
V$ is given by
\[
\alpha'(x) + \alpha^2(x) = 2[\tilde V(x)-\epsilon] \quad \Rightarrow
\quad \tilde V(x) = V(x) + \alpha'(x)
\]
The links of (\ref{intertwining}) with the traditional factorization
method \cite{Schro,Infe} are due to the identities
\be
H - \epsilon = A^{\dagger} A, \qquad \tilde H - \epsilon = A A^{\dagger}
\label{factor}
\ee
(see, e.g., Andrianov \cite{Andr}, Nieto \cite{MMN}, Sukumar \cite{Suku}). 
The Darboux transformation (\ref{intertwining}), in general, leads to a
new class of spectral problems of form (\ref{scr2})  with $\tilde V(x)$
essentially different from $V(x)$. Yet, the method allows some intriguing
exceptions. 

As recently discussed \cite{Dune,ks99,formin}, some special periodic
potentials $V(x) \equiv V(x+T)$ admit the Darboux transformations which
consist simply in coordinate displacements, $\tilde V(x) = V(x+\delta)$,
where $\delta =T/2$. As subsequently shown, the half period is not the
only displacement available. For a subclass of periodic Lam\'e potentials
\cite{Arscott}, we have detected the supersymmetric displacements $\delta$
varying continuously in the interval $(0,T)$ \cite{FMRS}. We propose to
call this phenomenon the {\it translational invariance with respect to
Darboux transformations}, or simply the {\it Darboux invariance}
\cite{FMRS}. Since very little is changed in $V(x)$ by just shifting the
argument, the Darboux displacements might look as `frustrated cases' of
the Darboux method (at least as far as the quest for new exact solutions
in quantum mechanics is concerned).  Yet, we shall show that the effect is
mathematically nontrivial and of tentative physical interest. We shall
formulate the necessary and sufficient conditions for a potential to be
Darboux invariant and find an intimate relation between the Darboux
invariance and elliptic functions.  Moreover, it turns out that the
Schr\"odinger equation with a Darboux invariant potential may be
integrated by quadratures for any value of $E$. This gives a convenient
tool to construct essentially new solvable potentials by applying the
finite difference B\"acklund algorithm \cite{Adler,FMH,MNR00}. 

\section{The Weierstrass condition}

Following our former observations \cite{FMRS}, we shall look for some
deeper reasons of the Darboux invariance. We shall thus consider a
hypothetical potential $V(x)$ (periodical or not)  which admits Darboux
transformations consisting in pure $x$-displacements. Let $A$ and
$A^{\dagger}$ be the first order differential operators
\be
A = \frac{1}{\sqrt{2}}\left[\frac{d}{dx} + \alpha(x, \delta)\right],
\qquad
A^{\dagger} =
\frac{1}{\sqrt{2}}\left[-\frac{d}{dx} + \alpha(x,\delta)\right]
\label{dintertwiners}
\ee
producing this effect, {\it i.e.\/},
\be
A H = H_{\delta} A
\label{dfactor}
\ee
where
\[
H_{\delta} = -\frac{1}{2} \frac{d^2}{dx^2} + V(x + \delta)
\]
Notice that if $\delta$ is one of the admissible Darboux displacement
(\ref{dintertwiners}-\ref{dfactor}), so is $-\delta$ (the formal
conjugation of (\ref{dintertwiners}) and the change of variable
$x\rightarrow x-\delta$ implying that $\alpha (x, -\delta) = -\alpha (x -
\delta, \delta)$ is the corresponding superpotential).  Note also that if
the Hamiltonian $H$ admits a Darboux displacement $\delta$
(\ref{dintertwiners}-\ref{dfactor}), then any displaced version
$H_{\delta'}$ of $H$ can be as well $\delta$-displaced
\[
A' H_{\delta'} = H_{\delta' + \delta} A', \qquad A' \equiv
\frac{1}{\sqrt{2}}\left[\frac{d}{dx} + \alpha(x+\delta', \delta)\right]
\]
The set ${\cal D}$ of all Darboux displacements $\delta$ generated by the
first order intertwiners (\ref{intertwiners}) for a given Hamiltonian $H$
is now a decisive element. Our previous study \cite{FMRS} shows that for
the one soliton or Lam\'e potentials with $n=1$ the allowed displacement
can be any $\delta \in (0,T)$ (where $T$ is either the real Lam\'e period
or $T=+\infty$ in the 1-soliton case). We shall see that even the
existence of a finite number of Darboux displacements can be a tight
structural information. Indeed, one has: 
\begin{itemize}
\item[]{\bf Proposition 1}. If the Hamiltonian $H$ admits three Darboux
displacements $\delta_1$, $\delta_2$, $\delta_3$, such that $\delta_1 +
\delta_2 +\delta_ 3 = 0$, then up to an additive constant, the potential
$V$ reduces to one of the Weierstrass functions. 
\end{itemize}

{\bf Proof} follows immediately from a sequence of mathematical results
\cite{Novi,Vese,Pere}, concerning the invariance of the Schr\"odinger's
Hamiltonians under the generalized Darboux transformations, where the
intertwiners $A$ in (\ref{intertwining})  can be differential operators of
arbitrary order.  Indeed, suppose $A_1$, $A_2$ and $A_3$ are three first
order Darboux operators $A_i= [d/dx + \alpha_i(x)]/\sqrt{2}$ inducing the
subsequent displacements
\be
H
\buildrel  \over \longrightarrow
H_{\delta_1}
\buildrel  \over \longrightarrow
H_{\delta_1 + \delta_2}
\buildrel  \over \longrightarrow
H_{\delta_1 + \delta_2 + \delta_3} = H. \label{3displ}
\ee
Then the product $D_3 = A_3 A_2 A_1$ must commute with the Hamiltonian
(\ref{scr1}), implying that $H$ and $D_3$ form a commuting Lax pair
(compare \cite{Lax,Novi}). Hence, there exists a constant $V_0\in{\mathbb
R}$ such that $\phi(x) = V(x) - V_0$ must fulfill the stationary KdV
equation (see also \cite{Pere}), leading to the 1-st order Weierstrass
equation \cite{Bateman}
\be
(\phi')^2 = 4 \phi^3 -g_2 \phi -g_3, \qquad g_2, g_3 = {\rm const}
\label{weiers1}
\ee
\hfill$\square$

The equation (\ref{weiers1}) admits two families of real solutions: 

$\bullet$ The {\it singular family} ({\bf S}) is given by the traditional
Weierstrass functions:
\be
\int_{-\infty}^\phi \frac{d \nu}{\sqrt{4 \nu^3 -g_2 \nu -g_3}} = x -a
\qquad \Rightarrow \qquad \phi(x) \equiv \wp (x-a; g_2, g_3)
\label{weiers2}
\ee
If $a,g_2,g_3$ are real, $\phi(x)$ is real too, but the family admits also
an analytic continuation to complex $a$. In fact, if $\omega'=i\tau$
($\tau\in{\mathbb R}$) is half-imaginary period of $\wp$, then one sees: 
$\wp (x - i\tau; g_2, g_3)^* = \wp(x + i \tau;g_2, g_3) = \wp (x - i \tau;
g_2, g_3)$. Thus, (\ref{weiers2})  defines as well

$\bullet$ The {\it regular family} ({\bf R}) is given as the `parallel
real section' of (\ref{weiers2}) for $a \rightarrow a + i\tau$
\be
\phi(x) \equiv \wp (x - a  - i \tau; g_2,g_3)
\label{weiers3}
\ee

To obtain a geometric image of both branches, the {\it phase portrait} of
(\ref{weiers1}) is relevant \cite{Poincare}. Take for simplicity $a=0$.
Interpreting $\phi$ and $\phi'$ respectively as the coordinate and
momentum of a hypothetical point particle, with $x$ meaning the `time',
one can view (\ref{weiers1}) as a dynamical law defining the `momentum' $p
=\phi'$ as a function of the `position' $\phi$, thus allowing $\phi$ to
move only in the permitted areas where $P(\phi)\equiv 4\phi^3 -g_2 \phi
-g_3 \geq 0$ and (\ref{weiers1}) is consistent with $(\phi')^2 \geq 0$. If
all three roots $e_1$, $e_2$, $e_3$ of $P(\phi)$ are real, $e_3 < e_2 \leq
e_1$, there are two allowed intervals $[R]=[e_3, e_2]$ and
$[S]=[e_1,+\infty)$ where $P(\phi)=(\phi')^2$ permits the real $\phi'$. 
The motions in $[S]$ typically depart from and return to the infinity at a
finite time $T$ (a real period of $\wp$);  their repetitions paint an
image of the periodic, singular Weierstrass functions. In turn, the
motions in $[R]$, in general, oscillate between two turning points $\phi_3
= e_3$, $\phi_2=e_2$, yielding the real, regular, bounded solutions of
(\ref{weiers1}) with a real period $T$.  If $e_1=(2-m)/3$, $e_2=(2m-1)/3$,
$e_3=-(m+1)/3$ and the oscillation period in $[R]$ is $T=2\omega$ (we
adopt the notation of \cite{Abram}), then we obtain the Lam\'e function
$\phi(x)=m{\rm sn}^2(x\vert m) - (m+1)/3$, but if $e_1 = e_2= 1/3 >e_3
=-2/3$ the oscillation time in $[R]$ tends to infinity and the motion
reproduces the one-soliton transparent well. 

As we have already observed, the regular solutions in $[R]$ can be as well
obtained by an analytic continuation of the singular Weierstrass
solutions in $[S]$. Thus, {\it e.g.\/}, the Lam\'e function ${\rm
sn}^2 x$ is given in terms of the regular branch of (\ref{weiers1}) by
\be
m{\rm sn}^2(x\vert m)= \wp (x+i\tau;g_2,g_3) + (m+1)/3 \label{lame}
\ee
where $g_2 = 4(m^2-m+1)/3, \ g_3=4(m-2)(2m-1)(m+1)/27$ (compare
\cite{Abram}) while the one soliton well is the $i\tau$-displaced case of
the singular solution $\wp (x)$ (see (31) in \cite{MNR00}). 

So far, (\ref{weiers1}-\ref{weiers2}) are just a necessary condition for
the existence of any 3-order Darboux symmetry of the initial Hamiltonian
$H$ (and by the same for the existence of a triple Darboux displacement
(\ref{3displ}) closing to identity). Quite remarkably, the condition turns
also sufficient, though the proof of this last fact is less evident.  We
shall therefore formulate an independent criterion which is both necessary
and sufficient for the existence of the Darboux displacements.

\section{The supersymmetric addition law}

Notice that even the existence of a single 1-st order intertwining
operator producing a displacement $\delta$ imposes strong restrictions on
the corresponding potential $V(x)$.  Of course, if $V$ is periodic, with a
real period $T$ and $V \neq {\rm const}$ then $\delta \neq nT$ ($n \in
{\mathbb Z}$).  Indeed, if $\delta =nT$, there would be a Darboux operator
(\ref{dintertwiners})  generating the identity transformation
$H_{\delta}=H$, {\it i.e.\/}, commuting with $H$, which is impossible
except if $V(x) \equiv {\rm const}$.  Assume now that (\ref{intertwiners}) 
is one of operators generating a Darboux displacement $\delta$ for the
Hamiltonian (\ref{scr1}); hence
\bea
\label{riccati1a}
-\alpha'(x) + \alpha^2(x) = 2[V(x)-\epsilon],\\[2ex]
\alpha'(x) + \alpha^2(x) = 2[V(x+\delta)-\epsilon]
\label{riccati2}
\eea
where $\epsilon$ is a factorization constant. Due to
(\ref{riccati1a}-{\ref{riccati2})
\bea
\label{pot1}
\alpha^2(x)  = V(x) + V(x+\delta) -2\epsilon,\\[2ex]
\alpha'(x) = V(x+\delta) - V(x)
\label{pot2}
\eea
Determining $\alpha(x)$ from (\ref{pot1}) one finds
\be
\alpha(x) = \pm \sqrt{V(x) + V(x+\delta) -2\epsilon}
\label{superpotencial}
\ee
Differentiating (\ref{superpotencial}) and comparing with
(\ref{pot2}) one thus arrives at the following functional equation:
\be
V(x) + V(x+ \delta) -\frac{1}{4} \left[ \frac{V'(x) + V'(x+
\delta)}{V(x) - V(x+ \delta)} \right]^2 = 2\epsilon 
\label{diferencias}
\ee
The eq. (\ref{diferencias}) turns out a necessary condition for $V(x)$ to
admit the Darboux displacement $V(x)  \rightarrow V(x+\delta)$ with the
factorization constant $\epsilon$. Inversely, suppose $V(x)$ fulfills
(\ref{diferencias})  with certain constants $\delta$ and $\epsilon$. Then
define $\alpha(x)$ by (\ref{superpotencial}), assuring automatically
(\ref{pot1}). Differentiating (\ref{superpotencial}) one obtains: 
\[
\alpha'(x) = \pm \frac12 \frac{V'(x) + V'(x+\delta)}{\sqrt{V(x) +
V(x+\delta) -2\epsilon}}
\]
If the sign in (\ref{superpotencial}) is ``$+$'' the superpotential
$\alpha(x)$ generates the Darboux displacement $V(x) \rightarrow V(x+
\delta)$, while the sign ``$-$'' yields the inverse displacement $V(x+
\delta) \rightarrow V(x)$. By choosing the proper sign $+$ and by applying
(\ref{diferencias}) one recovers (\ref{pot2}).  We thus arrived at

\begin{itemize}
\item[]{\bf Theorem 1}. The necessary and sufficient condition for $V(x)$
to admit a non-trivial Darboux displacement $\delta$ is that the left hand
side of (\ref{diferencias}) is independent of $x$. Its value defines the
factorization constant $\epsilon$ for the corresponding superpotential
$\alpha(x)$.
\end{itemize}

\noindent In order to admit a set ${\cal D} \subset {\mathbb R}$ of many
Darboux displacements, the potential $V(x)$ must satisfy a family of many
simultaneous difference-differential equations of type
(\ref{diferencias}). If ${\cal D}$ is non-trivial, the Proposition 1
implies that $V(x) = \phi(x) + V_0$ where $\phi$ is in the Weierstrass
class of functions. We shall see now that the set of conditions
(\ref{diferencias}) with continuous $\delta$ is indeed generic for the
Weierstrass functions. 

In fact, assume again $a=0$ in (\ref{weiers2}-\ref{weiers3}). Then,
examine the sense of (\ref{diferencias}) with $V(x)$ even, for $\delta
\neq nT$ (put $T=0$ if $V$ aperiodic). The constant $\epsilon$ has to
depend on $\delta$, $\epsilon=\epsilon(\delta)$. Denote for simplicity
${\cal E} (\delta)= -2\epsilon(\delta)$. Introducing the new variables
$u=x$ and $v=-\delta-x$, and using the fact that $V'(x)$ is odd, one can
write (\ref{diferencias}) in the form
\be
{\cal E} (u+v) + V(u) + V(v) =\frac{1}{4} \left[ \frac{V'(u)  -
V'(v)}{V(u) - V(v)} \right]^2
\label{ediferencias}
\ee
Notice now that this condition reduces to the well known addition formulae
for the Weierstrass singular (S) and regular (R) functions. Indeed, for
the singular branch (S) the traditional identity tells
\be
\wp(u+v) + \wp(u) + \wp(v) =\frac{1}{4} \left[ \frac{\wp'(u)  -
\wp'(v)}{\wp(u) - \wp(v)} \right]^2
\label{pdiferencias}
\ee
for all $u,v,u+v$ out of the singularities of (\ref{pdiferencias})  (see
e.g. Bateman \cite{Bateman}). Replacing now $u \rightarrow u-i \tau$, $v
\rightarrow v-i \tau$, $\wp(u-i \tau)= \phi(u)$, and making use of the
fact $\wp(u-2i \tau) \equiv \wp(u)$, one sees that for the regular branch
(R) 
\be
{\cal E}(u+v) + \phi(u) + \phi(v) =\frac{1}{4} \left[ \frac{\phi'(u)  -
\phi'(v)}{\phi(u) - \phi(v)} \right]^2
\label{vdiferencias}
\ee
where ${\cal E}(\delta) \equiv \wp(\delta)$ is the Weierstrass function of
the (S) branch linked with $\phi$ by analytic continuation. We thus have

\begin{itemize}
\item[]{\bf Theorem 2}. The addition laws (\ref{ediferencias}) for $V(x)$
and (\ref{pdiferencias},\ref{vdiferencias}) for the Weierstrass functions
are nothing else but the necessary and sufficient conditions for the
existence of a continuum of the Darboux displacements. 
\end{itemize}

\noindent Note, that we have thus detected a new sense of the traditional
addition formulae (\ref{pdiferencias}-\ref{vdiferencias}). Though these
formulae are a part of the textbook material on the elliptic functions
\cite{Bateman,Akhiezer}, the fact that they can be so simply obtained by
demanding the existence of the Darboux displacements
(\ref{riccati1a}-\ref{riccati2})  apparently, escaped attention. We
conclude that, without calling much attention, the Darboux displacements
were always present in the structure of the elliptic functions, explaining
the exact form of the addition laws. Some other points may be worth
making. 

{\it Observation 1.}
Though it was well established that the Weierstrass functions admit a 3-rd
order symmetry (leading to the 3-rd order stationary KdV, see
\cite{Novi,Vese,Pere}), as far as we know, it was not noticed that this
symmetry can be realized as a triple Darboux displacement.

{\it Observation 2.}
Though one knows that the algebraic addition laws limit the form of the
corresponding functions (permitting only rational or elliptic solutions),
it has not been noticed that the composition laws
(\ref{ediferencias}-\ref{vdiferencias}) have even stronger consequencies.
This is due to the fact that (\ref{ediferencias}-\ref{vdiferencias}) are
not purely algebraic, but have a form of difference-differential
equations. Of course, (\ref{pdiferencias},\ref{vdiferencias}) can be
rewritten as algebraic identities after eliminating $\phi'$ by using
(\ref{weiers1}), but this would reduce the implications to the traditional
Weierstrass theorem (stating that any meromorphic function $f$ which obeys
an algebraic addition law for $f(x), \ f(y)$ and $f(x+y)$, must be an
elliptic function;  see e.g. Akhiezer \cite{Akhiezer}, p.190). The
consequencies of the difference-differential laws
(\ref{ediferencias}-\ref{vdiferencias}) go beyond that. 

\begin{itemize}
\item[]{\bf Corollary} ({\it inverse addition theorem}). If a real,
differentiable, even function $\phi(x)$ fulfills the addition law
(\ref{vdiferencias}) for arbitrary $u, \ v,$ with ${\cal E}$ having an
isolated singularity at $0$, then $\phi(x)$ is one of Weierstrass
functions. 
\end{itemize}

{\bf Proof}. Indeed, if $\phi(x)$ is even and fulfills
(\ref{vdiferencias}), then it also satisfies (\ref{diferencias}) for any
$\delta=u+v$ whenever ${\cal E} (\delta)$ is finite. Then, due to our
Theorem~1, an arbitrary $\delta=u+v$ (out of singularities) belongs to the
admissible 1-susy displacements, and in view of the Proposition~1,
$\phi(x)$ belongs either to (R) or to (S) Weierstrass classes
(\ref{weiers1}-\ref{weiers2}). \hfill $\square$

Although our proof is immediate, the theorem (as far as we know) was never
proved, apparently since the supersymmetry methods has not been used in
the theory of the elliptic functions.

Let us also notice that the `supersymmetric sense' of the addition
formulae grants an explicit integrability of the Riccati equation
(\ref{riccati1}), making specially easy the use of the finite-difference
B\"acklund algorithm \cite{FMH,MNR00} to generalize the (R) or (S)
potentials. Indeed, for any $V(x)$ obeying (\ref{ediferencias}) the
special solutions $\alpha (x, \delta)$ of (\ref{riccati1}), generating the
displacements, are explicitly given by (\ref{superpotencial})  without
the need of solving any differential equation. Alternatively, using
(\ref{pot2}) one obtains:
\be
\alpha(x,\delta) =
\int [V(x+\delta) - V(x)]dx = \zeta(x) - \zeta(x+\delta) + \zeta(\delta),
\label{zeta}
\ee
where $\zeta(x)$ is the `non-elliptic' Weierstrass function \cite{Bateman}
and the last (constant) term in (\ref{zeta}) was determined consistently
with (\ref{superpotencial},\ref{ediferencias}). This might seem a limited
achievement (why to use the `supersymmetric machine' just to displace the
argument in $V(x)$?), but since $\epsilon(\delta)= \epsilon(-\delta)$, the
formula (\ref{zeta}) gives two independent solutions of the Riccati
equation (\ref{riccati1}) for the same factorization energy $\epsilon$.
Thus, the general solution can be easily obtained with the help of only
one quadrature (see e.g. \cite{cr99}), {\it i.e.\/}, in terms of a new
auxiliary function
\be
\tilde \alpha(x) = \Gamma e^{\int [ \zeta(x-\delta) -
\zeta(x+\delta)+2\zeta(\delta)]dx}= \Gamma
\frac{\sigma(x-\delta)}{\sigma(x+\delta)}e^{2\zeta(\delta)x}
\label{alphagral}
\ee
where $\sigma(x)$ is another non-elliptic Weierstrass function.  The
general solution of (\ref{riccati1}) then becomes:
\be
\alpha(x,\epsilon) = \frac{\alpha(x,\delta)-\alpha(x,-\delta)\tilde
\alpha(x)}{1-\tilde \alpha(x)} \label{gralalpha}
\ee

The possibility of solving generally the Riccati eq. (\ref{riccati1})  for
the Lame potential (\ref{lame}) is well known, but we have never seen an
argument so simple as the one based on the Darboux displacements. 
Moreover, by using (\ref{zeta}-\ref{alphagral}) with varying $|\delta|$,
one has explicit expressions for the superpotentials with different
factorization constants; a fact specially convenient for obtaining the
transformed Lam\'e potentials via the purely algebraic B\"acklund
algorithm \cite{FMH,MNR00}
\begin{eqnarray}
\hskip-0.9cm \alpha_2(x; \epsilon_1, \epsilon_2)
\!\!&\!\!= \!\!&\!\!
-\alpha_1(x;\epsilon_1) -
\frac{2(\epsilon_{1}-\epsilon_{2})}{
\alpha_1(x;\epsilon_1)- \alpha_1(x;\epsilon_2)}
\label{fractal}
\end{eqnarray}
where $\alpha_1$ can be either the displacement inducing solution
(\ref{zeta})  or the general one (\ref{gralalpha}). As an example, we have
used the general solution (\ref{zeta}-\ref{gralalpha})  to produce an
impurity of the Lam\'e potential inserting a bound state into the
lowest forbidden band (see Fig.1).

\begin{figure}
\centering \epsfig{file=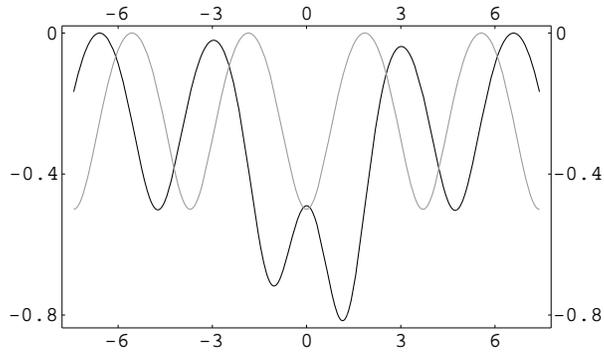, width=8cm}
\caption{\small The first order Darboux transformed potential (black
curve), defined by the general solution (\ref{gralalpha}) for the Lam\'e
function (gray curve)  with $\epsilon$ below the ground energy level.}
\end{figure}

An atypical application of (\ref{zeta}) permits to generate as well the
{\it complex Darboux displacements}. Indeed, it is known that the roots
$e_1$, $e_2$, $e_3$ of the Weierstrass polynomial $P(\phi)$ determine the
band edges of the nonsingular, periodic Weierstrass potentials
\cite{Bateman}. In particular, for the Lam\'e function (\ref{lame}) one
has $E_0=-\wp (\omega)/2 = -e_1/2=(m-2)/6$; $E_1=-\wp(\omega + i
\tau)/2=-e_2/2=(1-2m)/6$;  $E_{1'}=-\wp (i\tau)/2=-e_3/2 = (m+1)/6$. 
Hence, when applying the superpotential formula (\ref{zeta}) for
$\delta=i\tau+\kappa$, where $0\leq \kappa \leq \omega$, one obtains the
1-susy superpotentials with the factorization energies in the upper
forbidden band $[E_1,E_{1'}]$. The method can be very easily used to embed
two new bound states into the forbidden band by applying 2-nd order
Darboux \cite{s99} or B\"acklund algorithm (\ref{fractal}) (see Fig.2).
The higher order B\"acklund terms \cite{FMH,MNR00} can be as easily
constructed.

\begin{figure}[htbp]
\centering \epsfig{file=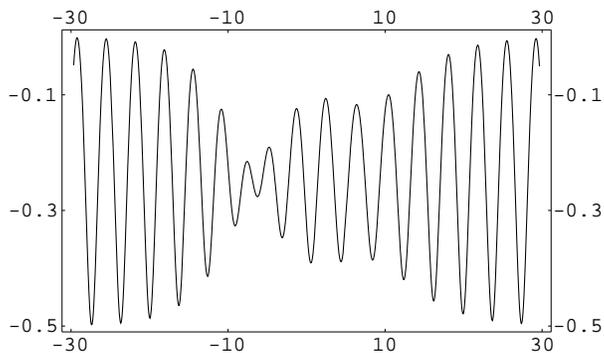, width=8cm}
\caption{\small The susy partner of (\ref{lame}) obtained by applying the
finite-difference B\"acklund algorithm (\ref{fractal})  with $\epsilon_1,
\ \epsilon_2$ inside of the spectral gap $[E_1,E_{1'}]$.}
\end{figure}

\vskip0.5cm
The authors acknowledge the support of CONACYT project 32086E (M\'exico). 
BFS is grateful for the support of the Russian Foundation for Basic
Research and for the kind hospitality at the Physics Department of
CINVESTAV in the spring of 2001. ORO acknowledges the support of
CINVESTAV, project JIRA'2001/17. 

\bigskip


\end{document}